\documentclass[aoas,nameyear,seceqn,dvips]{arximspdf}
\usepackage{graphics}
\doi{10.1214/09-AOAS294}
\volume{4}
\issue{2}
\pubyear{2010}
\firstpage{1081}
\lastpage{1104}

\makeatletter
\newtheorem{theorem}{Theorem}[section]
\newtheorem{lemma}{Lemma}
\newproclaim{remark}{Remark}
\makeatother

\begin{document}
\begin{frontmatter}

\title{Exact asymptotic distribution of change-point mle for change
in the mean of Gaussian sequences}
\runtitle{Exact asymptotic distribution of change-point mle}

\begin{aug}
\author[a]{\fnms{Stergios B.} \snm{Fotopoulos}\thanksref{a1}\ead[label=e1]{fotopo@wsu.edu}\corref{}},
\author[b]{\fnms{Venkata K.} \snm{Jandhyala}\thanksref{a1}\ead[label=e2]{jandhyala@wsu.edu}}\\
\and
\author[a]{\fnms{Elena} \snm{Khapalova}\ead[label=e3]{elena\_k@wsu.edu}}
\runauthor{S. B. Fotopoulos, V. K. Jandhyala and E. Khapalova}
\affiliation{Washington State University}
\address[a]{S. B. Fotopoulos\\E. Khapalova\\
Department of Management and Operations\\Washington State University\\
Pullman, Washington 99164-4736\\USA\\\printead{e1}\\
\phantom{E-mail: }\printead*{e3}} 
\address[b]{V. K. Jandhyala\\
Department of Statistics\\Washington State University\\
Pullman, Washington 99164-3113\\ USA\\\printead{e2}}
\thankstext{a1}{Supported by NSF Grant DMS-0806133.}
\end{aug}

\received{\smonth{12} \syear{2008}}
\revised{\smonth{9} \syear{2009}}

\begin{abstract}
We derive exact computable expressions for the asymptotic distribution
of the change-point mle when a change in the mean occurred at an unknown
point of a sequence of time-ordered independent Gaussian random
variables. The~derivation, which assumes that nuisance parameters such
as the amount of change and variance are known, is based on ladder
heights of Gaussian random walks hitting the half-line. We then show
that the exact distribution easily extends to the distribution of the
change-point mle when a change occurs in the mean vector of a
multivariate Gaussian process. We perform simulations to examine the
accuracy of the derived distribution when nuisance parameters have to be
estimated as well as robustness of the derived distribution to
deviations from Gaussianity. Through simulations, we also compare it
with the well-known conditional distribution of the mle, which may be
interpreted as a Bayesian solution to the change-point problem. Finally,
we apply the derived methodology to monthly averages of water discharges
of the Nacetinsky creek, Germany.
\end{abstract}


\end{frontmatter}

\section{Introduction}\label{sec1}

While modeling time-ordered data, one is concerned about the parameters
of the model being dynamically stable. One way of addressing the dynamic
instability of the model parameters is to model the time dependence of
parameters through a possible change at an unknown time-point so that
the parameters remain stable both before and after the unknown
change-point. Clearly, the methodology is extremely important from a
practical point of view, mainly because the changes in phenomena
observed over time usually occur unannounced, such as change in the
quality characteristic of a manufacturing process, changes in water or
air quality overtime, changes in the pattern of stock market indices and
so on. The~change-point problem allows modelers to detect the presence
of any such unknown change-points and further capture them through
either point or interval estimates. Such modeling has found applications
from all areas of scientific endeavor, including environmental
monitoring, global climatic changes, quality control, reliability,
financial and econometric time series, and medicine, to name a few. For
examples of real life applications, see Braun and M\"{u}ller (\citeyear{BrMu1998}) for
application of change-point methods in DNA segmentation and
bioinformatics; Fearnhead (\citeyear{Fe2006}), Ruggieri et al. (\citeyear{RuHeLaLa2009}) for applications
in geology; Perreault et al. (\citeyear{PeBeBoPa2000a}, \citeyear{PeBeBoPa2000b}) for application in hydrology;
Jaru\v{s}kov\'{a} (\citeyear{Ja1996}) for applications in meteorology; Fealy and
Sweeney (\citeyear{FeSw2005}) and DeGaetano (\citeyear{De2006}) for applications in climatology;
Kaplan and Shishkin (\citeyear{KaSh2000}) and Lebarbier (\citeyear{Le2005}) for applications in
signal processing; Andrews and Ploberger (\citeyear{AnPl1994}), and Hansen (\citeyear{Ha2000}) for applications in
econometrics; and Lai (\citeyear{La1995}), Wu, Cheng and Jeng (\citeyear{WuChJe2005}) and Zou, Qiu and Hawkins (\citeyear{ZoQiHa2009}) for
applications in statistical process control. Even though there are
recent advances in addressing multiple changes in scientific phenomena
[see Fearnhead (\citeyear{Fe2006}), Fearnhead and Liu (\citeyear{FeLi2007}), Gir\'{o}n, Moreno and Casella (\citeyear{GiMoCa2007})
and Seidou and Ouarda (\citeyear{SeOu2007})], the classical change-point literature is
most well developed in the case of a single unknown change-point in
time-ordered processes.

Classical change-point methods involve two fundamental inferential
problems, detection and estimation. Under the likelihood-based
approach, the detection part is addressed through likelihood ratio
statistics and their asymptotic sampling distributions. Maximum
likelihood estimation of an unknown change-point first begins with
obtaining the mle as a point estimate. Interval estimates of any desired
level, which are preferred over point estimates, can be constructed
around the mle, provided distribution theory for the mle is available.
However, distribution theory for a change-point mle can be analytically
intractable, particularly when no smoothness conditions are assumed
regarding the amount of change. In contrast, advances in the Bayesian
approach to change-point methodology have been occurring at a faster
pace. Ever since Markov chain Monte Carlo (MCMC) methods were seen as a
tool for overcoming the computational complexities in Bayesian analysis,
there has been rapid progress in the overall development of this
important methodological tool, and advances in Bayesian change-point
analysis have not lagged behind.

While the classical change-point problem dates back to Page (\citeyear{Pa1955}),
there has been a large amount of literature on the problem covering both
detection and estimation aspects. One may consult the monographs of
Brodsky and Darkhovsky (\citeyear{BrDa1993}, \citeyear{2000Brodsky}), Basseville and Nikiforov (\citeyear{BaNi1996}),
Cs\"{o}rg\H{o} and Horv\'{a}th (\citeyear{CsHo1997}), Chen and Gupta (\citeyear{ChGu2000}) and Wu
(\citeyear{Wu2005}), as well as a rich collection of references in these monographs
for a comprehensive account of various approaches to inference on
change-point problems. In reviewing the literature in terms of both
theory and applications, it becomes clear that the detection aspect of
the change-point problem attracted greater attention than its
counterpart of estimation. Perhaps this has not been accidental, in that
asymptotic theory for change-point estimators is technically a more
challenging problem than deriving asymptotic distribution theory for
change detection statistics. In an attempt to make estimation of the
unknown change-point more accessible to practitioners, the main purpose
of this paper is to derive exact computable expressions for the
asymptotic distribution of the maximum likelihood estimate (mle) of the
unknown change-point when a change occurs abruptly in the mean only of a
Gaussian process.

Asymptotic distribution theory for the change-point mle in the abrupt
case was first initiated by Hinkley (\citeyear{Hi1970}, \citeyear{Hi1971}, \citeyear{Hi1972}). While Hinkley
(\citeyear{Hi1970}) derived the asymptotic theory for the change-point mle in a
fairly general setup, the distribution was not in a computable form, and
was primarily technical in nature. It turns out that Hinkley (\citeyear{Hi1970})
computed the distribution for change in the mean of a normal
distribution only through certain approximations. While Hu and Rukhin
(\citeyear{HuRu1995}) provided a lower bound for the probability of the mle being in
error of capturing the true change-point, Jandhyala and Fotopoulos
(\citeyear{JaFo1999}) and Fotopoulos and Jandhyala (\citeyear{FoJa2001}) derived upper and lower
bounds and also suggested two approximations for the asymptotic
distribution of the change-point mle. Similarly, Borovkov (\citeyear{Bo1999}) also
provided only upper and lower bounds for the distribution of the
change-point mle. Thus, despite the attempts of various authors, the
problem of deriving computable expressions for the asymptotic
distribution of the change-point mle remained unsolved to date. It is
particularly striking that exact computable expressions for the
asymptotic distribution of the change-point mle have not been derived in
the literature for even selected distributions of the underlying process
such as the Gaussian and exponential distributions.

Tackling this important problem, we derive in this article exact
computable expression for the distribution of the change-point mle when
a change occurs in the mean only of a univariate or multivariate
Gaussian process. The~derived asymptotic distribution is not only exact
but is also quite elegant and can be computed in a simple and
straightforward manner. In fact, the result we derive demonstrates that
the second suggested approximation in Jandhyala and Fotopoulos (\citeyear{JaFo1999}) is
the exact solution to the problem, in the Gaussian case. It should be
pointed out that the distribution we derive assumes that the parameters
of the distribution before and after the change-point are known.
However, this should not pose difficulties, since Hinkley [(\citeyear{Hi1972}), page
520], in a theorem has shown that the asymptotic distribution of the
change-point mle remains the same even for unknown parameter scenarios.
From a practical point of view, this asymptotic equivalence result is
extremely important. In practice, apart from the change-point being
unknown, the parameters before and after the change-point also
invariably remain unknown. The~problem of deriving the distribution of
the change-point mle when the parameters are unknown is the one that
practitioners would be most interested, as opposed to the distribution
of the change-point mle for the case when the parameters are known.
There is no a priori reason to believe that the distributions of the
change-point mle for the known and unknown cases be asymptotically
equivalent. It is in this sense that the asymptotic equivalence result
of Hinkley (\citeyear{Hi1972}) plays a key role for practitioners. One only needs to
examine whether this asymptotic property holds well for reasonable
sample sizes, and for this we carry out a simulation study in Section~\ref{sec4}.

Since the exact solution derived in the paper assumes Gaussianity, it is
tempting to explore robustness of this exact computable expression when
the true process deviates from Gaussianity. If the derived result is
indeed robust to such departures, then it can be applied more widely
than merely Gaussian processes. While a simulation study covering a wide
class of non-Gaussian families of distributions may be of interest for
practitioners, in this paper we pursue a limited robustness study by
performing large scale simulations wherein the error process is assumed
to be symmetric and follows the $t$-distribution, or asymmetric and
follows the standardized chi-square distribution. In both cases, we
change the degrees of freedom from being small to large, so that one
approaches Gaussianity as the degrees of freedom become large.

Hinkley's approach to deriving distribution of the change-point mle is
perceived as the unconditional approach in the literature. Against this,
Cobb (\citeyear{Co1978}) proposed a conditional approach to the distribution of the
change-point mle, wherein the distribution of the mle is derived by
conditioning upon sufficient information on either side of the unknown
change-point. Since the exact distribution of the unconditional mle is
now available, it is relevant to compare the conditional and
unconditional distributions in terms of their performance, including
robustness properties. Thus, we have also included Cobb's conditional
distribution in our simulations. As pointed out by Cobb (\citeyear{Co1978}), since
the conditional distribution of the change-point mle can also be
interpreted as the Bayesian posterior for the change-point under a
uniform prior on the unknown change-point, the comparisons between the
two distributions have a broader appeal than what might appear at first
glance.

Finally, we apply the methodology derived in the paper to multivariate
analysis of hydrological data. The~data, previously analyzed in a
univariate setup by Gombay and Horv\'{a}th (\citeyear{GoHo1997}), represents averages
of log transformed water discharges for the Nacetinsky creek for the
months of February, July and August during the years 1951--1990. The~bivariate and trivariate change-point analysis shows that a significant
increase has occurred in the water discharges, whereas the univariate
change-point analyses show no significant changes in the mean water
flows.

The~organization of the paper is as follows. In Section \ref{sec2} we present
some general background regarding the change-point mle and its
asymptotic distribution. Then, we state the main theorem in Section \ref{sec3},
and the proof of the theorem is presented in Appendix \hyperref[appendA]{A}. While Section \ref{sec4}
consists of empirical assessment of the performance of derived theory
for the case of known and unknown parameters, Section \ref{sec5} contains the
multivariate change-point analysis of the Nacetinsky creek data.
Finally, Section \ref{sec6} concludes the paper with a discussion.

\section{Distribution of the mle} \label{sec2}

Let $Y_{1}, Y_{2}, \ldots,Y_{n}$, $n \ge 1$, be a sequence of
real-valued independent time ordered random variables defined on a
probability space $( \Omega,F,P )$. Let there be a natural number
$\tau_{n} \in\{ 1,2, \ldots,n - 1 \}$ such that $Y_{1}, Y_{2},
\ldots,Y_{\tau _{n}}$ have a common distribution $F_{1}$,
whereas the subsequent observations $Y_{\tau _{n} + 1}, Y_{\tau _{n} +
2}, \ldots,Y_{n}$ have a common distribution $F_{2}$ with $F_{1}
\ne F_{2}$. Here, the change-point $\tau_{n}$ is an unknown parameter
and should be estimated. The~likelihood function of $\tau_{n}$ is given
by $p_{n}( Y;\tau_{n} ) = \prod_{i = 1}^{\tau _{n}}f_{1}( Y_{i} )\prod_{i
= \tau _{n} + 1}^{n}f_{2}( Y_{i} )$, where the functions $f_{1}$
and $f_{2}$ are densities of $F_{1}$ and $F_{2}$, respectively, with
respect to some dominating measure $\mu( F_{1}, F_{2} \ll \mu)$. In the
sequel we assume that the densities $f_{1}$ and $f_{2}$ are known,
perhaps through known parameters. Following Hinkley (\citeyear{Hi1970}), the mle
$\hat{\tau} _{n}$ may be expressed as
\begin{equation}
\label{eq2.1}
\hat{\tau} _{n} = \mathop{\operatorname{arg\,max}}_{1 \le j \le n - 1}\sum_{i = 1}^{j} a(
Y_{i} ),
\end{equation}
where $a( Y_{i} ) = \log\{ f_{1}( Y_{i} ) / f_{2}( Y_{i} ) \},i = 1,
\ldots,n - 1$. For establishing distribution theory, it is convenient to
work with $\hat{\tau} _{n} - \tau_{n} \in\{ - \tau_{n} + 1, \ldots,n -
\tau_{n} - 1 \}$ instead of $\hat{\tau} _{n}$. Hence, we have
\begin{equation}
\xi_{n} = \hat{\tau} _{n} - \tau_{n}= \mathop{\operatorname{arg\,max}}_{ - \tau _{n} + 1 \le j
\le n - \tau _{n} - 1}\sum_{i = 1}^{\tau _{n} + j} a( Y_{i} ),
\end{equation}
where the maximizer is a result of the following two-sided random
walk $\Gamma( \cdot)$:
\begin{equation}
\label{eq2.3}
\qquad \Gamma_{n}( j;\tau_{n} ) = \cases{
\displaystyle \sum_{i =1}^{j} a( Y_{i}^{*} ) = \sum_{i = 1}^{j} X_{i}^{ *} = S_{j}^{
*},&\quad     $j \in\{ 1, \ldots,n - \tau_{n} - 1 \}$,\vspace*{2pt} \cr
 0, &\quad $j = 0$,\vspace*{2pt} \cr
 \displaystyle - \sum_{i = 1}^{ - j} a( Y_{i} ) = \sum_{i = 1}^{
- j} X_{i} = S_{ - j},&\quad    $j \in\{ - 1, \ldots, - \tau_{n} +
1 \} $.}
\end{equation}
Here, $\{ Y,Y_{i}\dvtx i \ge 1 \}$ and $\{ Y^{ *},Y_{i}^{ *} \dvtx i \ge 1 \}$ are
two independent sequences with independent and identical copies on $(
\mathbf{R},\mathrm{R} )$ such that $Y$ is distributed according
to $F_{1}$, and $Y^{ *}$ is distributed according to $F_{2}$. Note that
$X$ and $X^{ *}$ are real valued random variables defined
on $\mathbf{R}$. Also note that when $F_{1} \ne F_{2}$,
\begin{eqnarray}\label{eq2.4}
E( X ) &=& - \int_{ S} \log\{ f_{1}( x ) / f_{2}( x ) \}f_{1}( x )\mu(
dx ) = - K( f_{1},f_{2} )\nonumber\\
& =& - E_{f_{1}}\{ a( Y ) \} < 0\quad
\mbox{and}\nonumber\\[-8pt]\\[-8pt]
E( X^{ *} ) &=& \int_{ S} \log\{ f_{1}( x ) /
f_{2}( x ) \}f_{2}( x )\mu( dx
) = - K( f_{2},f_{1} )\nonumber\\
& =& E_{f_{2}}\{ a( Y^{
*} ) \} < 0,\nonumber
\end{eqnarray}
where $K$ is the usual Kullback--Leibler information. It can be seen that
(\ref{eq2.4}) is also related to the entropy function, which in many instances
is used for measuring the distinctness of probabilities. We assume
that $P( X > 0 ) > 0$. For $\theta > 0$, let
\begin{equation}\label{eq2.5}
\phi( \theta) = E\{ \exp( \theta X ) \}\quad \mbox{and}\quad \psi( \theta) = E\{ \exp(
\theta X^{ *} ) \}.
\end{equation}
Note that $\phi( \theta) = \psi( 1 - \theta)$. Moreover, $\phi( \theta)
\le 1, \forall\theta\in[ 0,1 ]$, since
\begin{eqnarray}\label{eq2.6}
\phi( \lambda) &=& \int_{ S} f_{1}( x )\{
f_{1}( x ) / f_{2}( x ) \}^{ - \lambda}
\mu(dx) = \int_{ S} f_{1}^{1 - \lambda} ( x
)f_{2}^{\lambda} ( x )\mu(dx)\nonumber\\[-8pt]\\[-8pt]
&\le&\biggl\{ \int_{ S} f_{1}( x )\mu(dx)
\biggr\}^{1 - \theta} \biggl\{ \int_{ S} f_{2}( x )\mu(dx) \biggr\}^{\theta} = 1.\nonumber
\end{eqnarray}

It is known that when $E( X ) < 0$, $P( X > 0 ) > 0$ and $\vartheta =
\sup\{ \theta > 0\dvtx \phi( \theta) \le 1 \}$, the asymptotic behavior of
the tail for the ultimate maximum, $M = \sup\{ S_{n}\dvtx n \in\mathbf{N}
\}$, can be described by the following three cases:
\begin{longlist}[(iii)]
\item[(i)] $\vartheta = 0$, the tail has a polynomial form (sub-exponential
case),

\item[(ii)] $\vartheta > 0$ and $\phi( \vartheta) < 1$ an intermediate case,

\item[(iii)] $\vartheta > 0$ and $\phi( \vartheta) = 1$ the Cram\'{e}r's case.
\end{longlist}

Now, in a sequence of observations for which $F_{1} \ne F_{2}$, the $\mu$-derivatives also satisfy $f_{1} \ne f_{2}$. From (\ref{eq2.6}), it is clear
that the choice of $\vartheta$ greater than zero for which (iii) is
satisfied is $\vartheta = 1$, the unity. Consequently, it follows that
$X$ satisfies Cram\'{e}r's condition. Furthermore, merely noting
that $\psi( \vartheta) = \phi( 1 - \vartheta)$, it follows that $X^{ *}$
also satisfies Cram\'{e}r's condition. This observation implies
that \mbox{$\vartheta = \vartheta^{ *} = 1$}, in Proposition~1 of Jandhyala and
Fotopoulos (\citeyear{JaFo1999}) for general distributions including Gaussian random
variables.

It also follows that $\phi( \theta) < 1,\forall\theta\in( 0,1 )$ and that
$\phi$ is strictly convex on \mbox{$\theta\in( 0,1 )$}. This suggests that
$\phi( \theta)$ attains its minimum at a unique $\theta_{0} \in( 0,1 )$
such that $\phi( \theta_{0} ) = \inf_{\theta \in ( 0,1 )}\phi( \theta) <
1$. This firmly establishes that assumptions~\mbox{1--3} in Jandhyala and
Fotopoulos (\citeyear{JaFo1999}) are no more required and that they hold naturally
whenever $F_{1} \ne F_{2}$, and $P( X > 0 ) > 0$ are satisfied.

In this paper we are interested in deriving the distribution of the
limiting variable~$\xi_{\infty}$, by letting $n \to\infty$ in such a way
that $\tau_{n} \to\infty$ and $n - \tau_{n} \to\infty$. In this regard,
it has been shown that $\xi_{\infty}$ is a proper random variable and
$\xi_{n} \to\xi_{\infty}$ a.s. [see, e.g., Fotopoulos and Jandhyala (\citeyear{FoJa2001})].

We begin by stating a theorem found in Fotopoulos (\citeyear{Fo2009}). For all
purposes, this result is a restatement of Theorem 2 in Jandhyala and
Fotopoulos (\citeyear{JaFo1999}).

\begin{theorem}\label{teo2.1}
 Let $F_{1} \ne F_{2}$ and $P( X > 0 ) > 0$. Then, the
probability distribution of $\xi_{\infty}$ is given by
\[
P( \xi_{\infty} = j ) =
\cases{\displaystyle
P( T_{1}^{ +} = \infty) \biggl\{P( T_{1}^{ -} > - j )\vspace*{2pt}\cr
\displaystyle\qquad\hspace*{40pt}  - \int_{ 0 +} ^{ \infty} P( M^{ *} \ge x )P(T_{1}^{ -} > - j \cap S_{ - j} \in dx ) \biggr\},\vspace*{2pt}\cr
  \displaystyle\qquad j \le - 1, - 2, \ldots,\vspace*{2pt}\cr
 \displaystyle P( T_{1}^{ +} = \infty) P( T_{1}^{* +} = \infty),\qquad  j = 0,\vspace*{2pt}\cr
  P( T_{1}^{* +} = \infty) \biggl\{ P( T_{1}^{ * -} > j )\vspace*{2pt}\cr
  \displaystyle\hspace*{45pt}\qquad  - \int_{ 0 +} ^{\infty} P( M \ge x )P( T_{1}^{ * -} > j \cap S_{j}^{ *} \in dx )\biggr\},\vspace*{2pt}\cr
  \qquad   j = 1, 2,  \ldots,}
\]
where $T_{1}^{ +} : =
\inf\{ j > 0\dvtx S_{j} > 0 \}, T_{1}^{ -} : = \inf\{ j > 0\dvtx S_{j} \le 0 \}$
and $M: = \max_{0 \le n}S_{n}$, and $M^{ *}, T_{1}^{ * +}$
and $T_{1}^{ * -}$ are defined in a similar manner.
\end{theorem}

The~convergence rate of the above asymptotic result is of interest for
purposes of both theory and practice. Knowledge about the convergence
rate allows one to judge the appropriateness of the sample size and
other ancillary parameters for which the asymptotic distribution can be
utilized for finite sample sizes without committing disproportional
errors. In this regard, both Borovkov (\citeyear{Bo1999}) and Jandhyala and
Fotopoulos (\citeyear{JaFo2000}) derived important results that establish the
convergence rate applicable to Theorem \ref{teo2.1}. We state here some relevant
facts from these articles and then formulate a theorem without proof
that establishes a bound for the total variation distance between the
finite sample and infinite sample distributions of the change-point mle.

From Theorem 2 of Jandhyala and Fotopoulos (\citeyear{JaFo2000}), we have
\[
\sup_{B \in \mathrm{B}_{\tau _{n},n}}| P( \xi_{n} \in B ) - P(
\xi_{\infty} \in B ) |=P( \xi_{\infty} \le - \tau_{n}\mbox{ or
}\xi_{\infty} \ge n - \tau_{n} ),
\]
where $\mathsf{B}_{\tau _{n},n}$ is the Borel $\sigma $-field defined on
$\mathbf{Z}_{\tau _{n},n} \equiv\{ - \tau_{n} + 1, \ldots, 0, \ldots,\break n - \tau_{n} - 1 \}$. Then, as argued in Jandhyala and
Fotopoulos (\citeyear{JaFo2000}), upon augmenting $\mathsf{B}_{\tau _{n},n}$ into the
Borel $\sigma $-filed on $\mathbf{Z}$, it follows that the total
variation distance between $\xi_{n}$ and $\xi_{\infty}$ defined by
\[
d_{\mathrm{TV}}( \xi_{n},\xi_{\infty} ) = \sup_{B \in \mathsf{B}} |
P( \xi_{n} \in B ) - P( \xi_{\infty} \in B )
|
\]
may be seen to yield
\begin{equation}\label{eq2.7}
d_{\mathrm{TV}}( \xi_{n},\xi_{\infty} ) =P( \xi_{\infty} \le - \tau_{n}\mbox{
or }\xi_{\infty} \ge n - \tau_{n} ).
\end{equation}
The~following theorem, which provides a bound for $d_{\mathrm{TV}}(
\xi_{n},\xi_{\infty} )$, follows immediately upon applying (\ref{eq2.7}) into
Theorem 1 of Borovkov (\citeyear{Bo1999}).

\begin{theorem}\label{teo2.2}
Let $F_{1} \ne F_{2}$ and $P( X > 0 ) > 0$. Let
 $\xi_{n}$ and  $\xi_{\infty}$  be the centered random
variables of the change-point mle for finite and infinite samples,
respectively. Then, the total variation distance between  $\xi_{n}$ and
 $\xi_{\infty}$ admits the inequality given by
\[
d_{\mathrm{TV}}( \xi_{n},\xi_{\infty} ) \le 4\max\{ \phi( \theta_{0} )^{\tau
_{n}}, \phi( \theta_{0} )^{n - \tau _{n}} \},
\]
where $\phi( \theta_{0} ) = \inf_{\theta \in ( 0,1 )}\phi(
\theta) < 1$.
\end{theorem}

Theorem \ref{teo2.2} clearly establishes a geometric rate of convergence as
$\xi_{n}$ approaches $\xi_{\infty}$, asymptotically. The~above result is
more friendly from a computational point of view than Theorem 3 of
Jandhyala and Fotopoulos (\citeyear{JaFo2000}).

While Theorem \ref{teo2.1} provides the probability distribution of $\xi_{\infty}$, the expressions therein are still only of technical interest. The~main
problem is that, as far as we know, a computable expression for the
distribution function $M( x )$ [or $M^{ *} ( x )]$ is not available in
the literature. Clearly, the behavior of $1 - M( x )$ (or $1 -
M^{ *} ( x ) )$ depends upon the characteristics of the underlying
distributions $f_{1}$ and $f_{2}$, in study. Moreover, the term $P(
T_{1}^{ +} = \infty)$ that appears in both Theorems \ref{teo2.1} and \ref{teo2.2} may also
be unavailable for computation unless we know the exact distribution of
$S_{n}$, for all $n \in N$. Thus, the determination of an exact
expression for the distribution of $M$ for any general distribution is
beyond analytical scope, and consequently, an exact computable form for
the probability distribution $P( \xi_{\infty} = j )$, $j \in\mathbf{Z}$, in
Theorem \ref{teo2.1} is also analytically not tractable. To this extent, in this
paper we shall concentrate on developing the analysis by assuming that
the underlying process is of Gaussian type.

\section{Asymptotic distribution of the mle under Gaussian
processes} \label{sec3}

We shall establish the main theorem regarding computationally accessible
distribution of~$\xi_{\infty}$ first under the univariate Gaussian case.
Subsequently, we shall illustrate how the univariate case itself can be
directly applied to the more general multivariate setup.

\subsection{The~univariate Gaussian case} \label{sec3.1}

We begin by assuming that the underlying process is univariate Gaussian,
and the means before and after the change-point are given by $\mu_{1},
\mu_{2}$, wherein we let $\mu_{1} \ne\mu_{2}$. We do assume that the
standard deviation $\sigma$ is known and remains the same throughout the
sampling period. Clearly, the likelihood ratios in (\ref{eq2.1}) may then be
expressed as
\begin{eqnarray}\label{eq3.1}
X &=& - a( Y ) = \log\{ f_{2}( Y ) /
f_{1}( Y ) \}\nonumber\\
& =& \log\biggl\{ \frac{1}{\sqrt{2\pi \sigma
^{2}}} e^{ - ( Y - \mu _{2} )^{2} / 2\sigma ^{2}} \Big/
\frac{1}{\sqrt{2\pi \sigma ^{2}}} e^{ - ( Y - \mu _{1} )^{2}
/ 2\sigma ^{2}} \biggr\}\\
&\phantom{;}= _{\mathsf{D}}& - \frac{( \mu _{1} - \mu _{2} )^{2}}{2\sigma ^{2}}
- \frac{( \mu _{1} - \mu _{2} )}{\sigma} Z,\nonumber
\end{eqnarray}
where $Z\sim N( 0,1 )$, and, similarly,
\begin{equation}\label{eq3.2}
X^{ *} = _{\mathsf{D}} - \frac{( \mu _{1} - \mu _{2} )^{2}}{2\sigma ^{2}} +
\frac{( \mu _{1} - \mu _{2} )}{\sigma} Z^{ *},
\end{equation}
where $Z^{ *} \sim N( 0,1 )$, and is independent of $Z$. Note that in
this case, the random variables $X$ and $X^{ *}$ are both identically
distributed with means $E( X ) =E( X^{*} ) = - \eta^{2} / 2 < 0$ and
variances ${\mathop{\operatorname{var}}} ( X ) ={\mathop{\operatorname{var}}} (
X^{*} ) =\eta^{2}$, where $\eta = \frac{| \mu _{1} - \mu _{2}
|}{\sigma}$ represents the standardized amount of change. Hence, it is
sufficient to confine our analysis to only one side of the random walk
$\Gamma( \cdot)$.

Under the formulation in (\ref{eq3.1}), it can be seen that $S_{n} = _{\mathsf{D}} -
n\frac{\eta ^{2}}{2} - \eta\sqrt{n} Z$, where again $Z\sim N( 0,1 )$.
Note [Asmussen (\citeyear{As1985}), Corollary 4.4] that when $E( X ) < 0$, the ladder
height distribution given by $G_{ +} (dx) = P( S_{T_{1}^{ +}}
\in dx \cap T_{1}^{ +} < \infty)$ is defective. Thus, $\| G_{ +}
\| = P( T_{1}^{ +} < \infty) < 1$ and $\frac{1}{E( T_{1}^{ -} )} =1 -
\| G_{ +} \| = P( T_{1}^{ +} = \infty) = P( M = 0 )$. We shall now state
our main theorem, which provides a computable expression for the
distribution of $\xi_{\infty}$. The~computability of the terms in the
expression will be demonstrated in the discussion following the theorem.
The~proof of the theorem is presented in Appendix \hyperref[appendA]{A}. Subsequent to the
theorem, we state a corollary, which establishes a closed form
computable expression for the bound in Theorem \ref{teo2.2}.

\begin{theorem}\label{teo3.1}
 Suppose that the time-ordered sequence $Y_{1},
Y_{2}, \ldots,Y_{n}$, \mbox{$n \ge 1$}, is such that $Y_{i}\sim N(
\mu_{1},\sigma^{2} ), i = 1, \ldots,\tau_{n}$, and $Y_{i}\sim
N( \mu_{2},\sigma^{2} )$,\break  $i = \tau_{n} + 1, \ldots,n$. Then, the
probability distribution of $\xi_{\infty}$ is given by
\[
P( \xi_{\infty} = k ) = \cases{
 ( 1 -\| G_{ +} \| ) \bigl( q_{| k |} - \|
G_{ +} \| \tilde{q}_{| k |} \bigr),&\quad     $k = \pm 1, \pm 2, \ldots,$\vspace*{2pt}\cr
 ( 1 - \| G_{ +} \| )^{2},&\quad           $k =0$,}
\]
where $1 - \| G_{ +} \|= \exp\{ - \sum_{j = 1}^{\infty}
\frac{1}{j}\bar{\Phi} ( \eta\sqrt{j} / 2 ) \}$ and $q_{k} = E\{ I(
T_{1}^{ -} > k ) \}$,
$\tilde{q}_{k} = E\{ e^{ - S_{k}}I( T_{1}^{ -} > k ) \}$, $k = 1, 2, \ldots$ and $q_{0} = \tilde{q}_{0} = 1$.
\end{theorem}

It is fairly straightforward to state the bound in Theorem \ref{teo2.2} for the
Gaussian case. Specifically, it follows that the total variation
distance in the Gaussian case admits
\begin{equation}
\label{eq3.3}
d_{\mathrm{TV}}( \xi_{n},\xi_{\infty} ) \le 4\max\biggl\{ \exp\biggl( -
\frac{\eta ^{2}\tau _{n}}{8} \biggr), \exp\biggl( - \frac{\eta ^{2}(
n - \tau _{n} )}{8} \biggr) \biggr\}.
\end{equation}

\subsection{The~multivariate Gaussian case} \label{sec3.2}

Here, we let $\{ Y,Y_{i}\dvtx i \in\mathbf{N} \}$ be a sequence of
time-ordered independent Gaussian elements defined on $\mathrm{R}^{d}$,
the $d$-dimensional Euclidean space with $f( x;\mu_{d \times
1},\Sigma_{d \times d} )$ denoting the corresponding
probability density function. In the sequel, mainly for convenience, we
represent the parameter only as $( \mu,\Sigma)$ by dropping the
respective dimension subscripts. Let the parameter $( \mu,\Sigma)$
change from its initial value of $( \mu_{1},\Sigma)$ to $(
\mu_{2},\Sigma)$, at some unknown index point $\tau_{n} \in\{ 1,2, \ldots,n - 1 \}$, with mean vectors $\mu_{1},\mu_{2} \in\Theta$, and
common variance-covariance matrix $\Sigma$. For reason of convenience, we
assume that $\Sigma$ is positive definite and the mean vectors satisfy
$\mu_{1} \ne\mu_{2}$.

The~functional $\langle x,y \rangle $ denotes the usual inner product and
the extended semi-norm is defined if there exists a covariance operator
$\Sigma$ such that $\| x \|_{\Sigma} ^{2} = \langle\Sigma x,x \rangle$.
Then, we may write $Y = _{\mathsf{D}} \mu_{1} + \Sigma^{1 / 2}\mathbf{Z}$ for
all data before the change-point, where~\textbf{Z} is a $d$-variate
standard normal vector. Consequently, the random variable $X = - \ln f(
Y;\mu_{1},\Sigma) / f( Y;\mu_{2},\Sigma)$ is expressed as
\begin{eqnarray}
X &=& \tfrac{1}{2}\{ \langle\Sigma^{ - 1}( Y - \mu_{1} ),Y - \mu_{1}
\rangle - \langle\Sigma^{ - 1}( Y - \mu_{2} ),Y - \mu_{2} \rangle\}\nonumber\\[-8pt]\\[-8pt]
&\phantom{;}=
_{\mathsf{D}}& - \tfrac{1}{2}\| \mu_{1} - \mu_{2} \|_{\Sigma ^{ - 1}}^{2} - \|
\mu_{1} - \mu_{2} \|_{\Sigma ^{ - 1}}Z,\nonumber
\end{eqnarray}
where $Z$ now stands for the standard normal random variable with mean
zero and variance one.

Similarly, for data after the change-point, we have $Y = _{\mathsf{D}} \mu_{2}
+ \Sigma^{1 / 2}\mathbf{Z}^{ *}$, where~$\mathbf{Z}^{ *}$ is the
$d$-variate standard normal vector, and in this case, we obtain
\begin{eqnarray}\label{eq3.5}
X^{*} &=& \ln f( Y;\mu_{1},\Sigma) / f( Y;\mu_{2},\Sigma)\nonumber\\[-8pt]\\[-8pt]
&\phantom{;}= _{\mathsf{D}}& -
\tfrac{1}{2}\| \mu_{1} - \mu_{2} \|_{\Sigma ^{ - 1}}^{2} + \| \mu_{1} -
\mu_{2} \|_{\Sigma ^{ - 1}}Z^{ *},\nonumber
\end{eqnarray}
where $Z^{ *}$ is univariate standard normal independent of $Z$. Upon
letting $\eta = \| \mu_{1} - \mu_{2} \|_{\Sigma ^{ - 1}}$ represent the
amount of standardized change in the means, it should be clear that the
multivariate case translates itself into a corresponding univariate case
with $\eta$ as defined above.

\section{Performance of the distribution of the change-point mle} \label{sec4}

In this section we wish to assess the performance of the derived
asymptotic distribution in two different ways. First, we investigate the
equivalence result of Hinkley (\citeyear{Hi1972}) and, second, we compare the derived
distribution of the mle with the conditional distribution of mle as
derived by Cobb (\citeyear{Co1978}).

\subsection{Distribution of the change-point mle for known and
unknown parameters} \label{sec4.1}

The~assumption of known parameters does not apply in practice, and it is
common that they must be estimated from the data. While Hinkley (\citeyear{Hi1972})
has shown asymptotic equivalence of change-point mle under both known
and estimated cases, its applicability to sample sizes of practical
interest requires empirical evidence. This issue is perhaps even more
important in the multivariate case, mainly because the multivariate case
involves estimation of many more parameters. As discussed in Sections \ref{sec2}
and \ref{sec3}, for comparing the closeness of two distributions, we find it
convenient to utilize the total variation distance measure, which for
discrete random variables $X$ and $Y$ is given by $d_{\mathrm{TV}}( X,Y ) =
\frac{1}{2}\sum_{i \in \mathbf{Z}} | P( X = i ) - P( Y = i ) |$.

Simulations are performed by letting the parameter choices for sample
size and true change-point be as follows: $n = 40$, $\tau = 20$; $n =
60$, $\tau = 20$; $n = 60$, $\tau = 30$; $n = 100$,
$\tau = 20$; $n = 100$, $\tau = 30$; $n = 100$, $\tau = 40$
and $n = 100$, $\tau = 50$. For each of the above cases, the
choice of values for $\eta$ are set at $\eta = 1.0, 1.5,
2.0, 2.5$. The~results for univariate and bivariate cases
based on 500,000 simulations for each individual scenario are presented
in Tables \ref{tab1} and \ref{tab2}, respectively. As one might expect, the situation
of known parameters yields excellent agreement with the theoretical
distribution in both tables, irrespective of the sample size as well as
the location of the change-point. When parameters are estimated, the
univariate case (Table \ref{tab1}) shows very good to extremely good agreement
with the theoretical distribution. The~values, for even the bivariate
case (Table \ref{tab2}), show very good agreement except when $\eta$ is very
small ($\eta = 1$).

\begin{table}
\caption{Total variation distances of known and estimated empirical distributions
(based on 500,000 simulations) from theoretical distribution of
change-point mle in the univariate case}
\label{tab1}
\begin{tabular*}{\textwidth}{@{\extracolsep{\fill}}lccccccccc@{}}
\hline
$\bolds{n}$ & \multicolumn{1}{c}{$\bolds{\tau}$} & \multicolumn{2}{c}{$\bolds{\eta = 1}$} & \multicolumn{2}{c}{$\bolds{\eta = 1.5}$} & \multicolumn{2}{c}{$\bolds{\eta = 2}$}
 & \multicolumn{2}{c@{}}{$\bolds{\eta = 2.5}$}\\[-6pt]
 &  & \multicolumn{2}{c}{\hrulefill} & \multicolumn{2}{c}{\hrulefill} & \multicolumn{2}{c}{\hrulefill} & \multicolumn{2}{c@{}}{\hrulefill}\\
 &  & \multicolumn{1}{c}{\textbf{Known}} & \textbf{Est.} & \textbf{Known} & \textbf{Est.} & \textbf{Known} & \textbf{Est.} & \textbf{Known} & \textbf{Est.}\\
\hline
100 & 20 & 0.0106 & 0.0665 & 0.0070 & 0.0264 & 0.0033 & 0.0139 & 0.0014 & 0.0082\\
100 & 30 & 0.0113 & 0.0493 & 0.0065 & 0.0205 & 0.0032 & 0.0104 & 0.0021 & 0.0057\\
100 & 40 & 0.0112 & 0.0437 & 0.0065 & 0.0189 & 0.0033 & 0.0091 & 0.0020 & 0.0050\\
100 & 50 & 0.0109 & 0.0412 & 0.0068 & 0.0176 & 0.0040 & 0.0082 & 0.0022 & 0.0044\\[3pt]
\phantom{0}60 & 20 & 0.0105 & 0.0721 & 0.0070 & 0.0298 & 0.0033 & 0.0155 & 0.0014 & 0.0086\\
\phantom{0}60 & 30 & 0.0112 & 0.0641 & 0.0065 & 0.0271 & 0.0032 & 0.0133 & 0.0021 & 0.0076\\[3pt]
\phantom{0}40 & 20 & 0.0104 & 0.0852 & 0.0070 & 0.0383 & 0.0033 & 0.0191 & 0.0014 & 0.0105\\
\hline
\end{tabular*}
\end{table}

\begin{table}[b]
\caption{Total variation distances of known and estimated empirical distributions
(based on 500,000 simulations) from theoretical distribution of
change-point mle in the bivariate case}
\label{tab2}
\begin{tabular*}{\textwidth}{@{\extracolsep{\fill}}lccccccccc@{}}
\hline
$\bolds{n}$ & \multicolumn{1}{c}{$\bolds{\tau}$} & \multicolumn{2}{c}{$\bolds{\eta = 1}$} & \multicolumn{2}{c}{$\bolds{\eta = 1.5}$} & \multicolumn{2}{c}{$\bolds{\eta = 2}$}
 & \multicolumn{2}{c@{}}{$\bolds{\eta = 2.5}$}\\[-6pt]
 &  & \multicolumn{2}{c}{\hrulefill} & \multicolumn{2}{c}{\hrulefill} & \multicolumn{2}{c}{\hrulefill} & \multicolumn{2}{c@{}}{\hrulefill}\\
 &  & \multicolumn{1}{c}{\textbf{Known}} & \textbf{Est.} & \textbf{Known} & \textbf{Est.} & \textbf{Known} & \textbf{Est.} & \textbf{Known} & \textbf{Est.}\\
\hline
100 & 20 & 0.0108 & 0.0991 & 0.0066 & 0.0376 & 0.0035 & 0.0197 & 0.0018 & 0.0126\\
100 & 30 & 0.0110 & 0.0718 & 0.0065 & 0.0281 & 0.0034 & 0.0153 & 0.0016 & 0.0099\\
100 & 40 & 0.0119 & 0.0624 & 0.0070 & 0.0252 & 0.0044 & 0.0135 & 0.0017 & 0.0075\\
100 & 50 & 0.0121 & 0.0595 & 0.0076 & 0.0236 & 0.0040 & 0.0126 & 0.0016 & 0.0075\\[3pt]
\phantom{0}60 & 20 & 0.0107 & 0.1140 & 0.0066 & 0.0466 & 0.0035 & 0.0248 & 0.0018 & 0.0157\\
\phantom{0}60 & 30 & 0.0107 & 0.1006 & 0.0065 & 0.0410 & 0.0034 & 0.0218 & 0.0016 & 0.0146\\[3pt]
\phantom{0}40 & 20 & 0.0105 & 0.1383 & 0.0065 & 0.0647 & 0.0035 & 0.0350 & 0.0018 & 0.0233\\
\hline
\end{tabular*}
\end{table}

\subsection{Unconditional change-point mle against Cobb's
conditional mle} \label{sec4.2}

Cobb (\citeyear{Co1978}) derived conditional distribution of the change-point mle by
conditioning upon sufficient observations around the true change-point,
which according to Cobb (\citeyear{Co1978}) is also equivalent to the Bayesian
posterior when the prior on the unknown change-point is uniform. If
$\delta$ denotes the number of data points to be considered on either
side of $\hat{\tau} _{n}$, then Cobb's conditional solution for $l \in\{
- \delta, \ldots,\delta\}$ is given by
\begin{eqnarray}
\label{eq4.1}
&&P( \hat{\tau} _{n} - \tau_{n} = l | Y_{\hat{\tau} _{n}
- \delta + 1}, \ldots,Y_{\hat{\tau} _{n} + \delta} )\nonumber\\[-8pt]\\[-8pt]
&&\qquad  \cong
p_{n}( Y;\hat{\tau} _{n} + l ) \Big/ \sum_{l = - \delta}
^{\delta} p_{n}( Y;\hat{\tau} _{n} + l ).\nonumber
\end{eqnarray}
The~method of choosing $\delta$ is clearly detailed in Cobb (\citeyear{Co1978}). It
is then relevant to compare the unconditional distribution of the mle
derived in Section \ref{sec3} with the above conditional solution. Also, we
investigate the robustness of the exact limiting distribution for
departures from normality through simulations, limiting the study to the
univariate framework only. Here, incorporating both symmetric and
asymmetric distributions, the error structures are modeled by the
standardized~$t_{\nu}$ and~$\chi_{\nu} ^{2}$ distributions.

\begin{figure}

\includegraphics{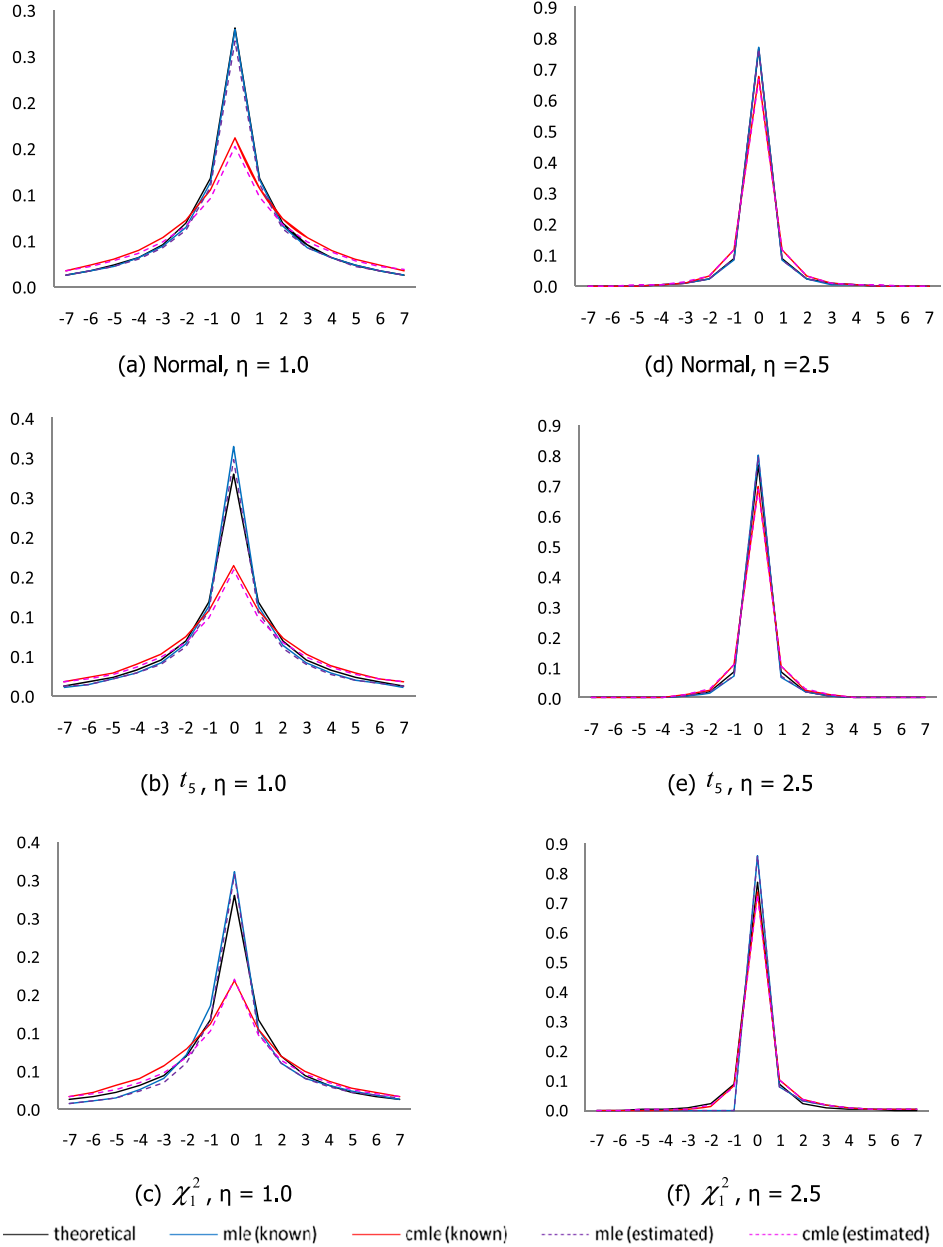}

\caption{Plots of theoretical mle, empirical mle (known), empirical mle
(estimated), empirical cmle (known) and empirical cmle (estimated)
distributions of the centered change-point when $n = 100, \tau = 50$
under normal \textup{(a)}; $t_{5}$ \textup{(b)} and
$\chi_{1}^{2}$ \textup{(c)} when $\eta = 1.0$;
and normal \textup{(d)}; $t_{5}$ \textup{(e)} and
$\chi_{1}^{2} $ \textup{(f)} when $\eta = 2.5$.}\label{fig1}
\end{figure}

For simplicity, we let only $\eta = 1.0$ and $\eta =  2.5$, and then
perform simulations for all the choices of sample sizes and true
change-points considered in Section~\ref{sec4.1}. The~choices of $\nu$ under
$t_{\nu} $-distribution were $\nu = 5, 10, 20$ and they were
$\nu = 1, 5, 20$ under $\chi_{\nu} ^{2}$-distribution. Note that while
implementing Cobb's conditional solution, we determined the value of
$\delta$ so that the error rate detailed in Cobb (\citeyear{Co1978}) is close to $10^{
- 5}$. To save space, we present the computed distributions (based on
50,000 simulations) in the form of figures only, and that too only for
the case of $n = 100, \tau = 50$. Figure \ref{fig1}(a--c)
correspond to the cases of normal, $t_{5}$ and $\chi_{1}^{2}$
distributions when $\eta = 1.0$, and Figure \ref{fig1}(d--f)
correspond to the same cases when $\eta = 2.5$.

For the remaining cases, we summarized the computed distributions
through Bias and mean square error (MSE), and to save space, we only
describe the salient features of these computations. It can be seen from
Figure \ref{fig1}(a) that in the normal case, the unconditional distributions
under both known and estimated cases are almost identical and they
closely agree with the theoretical distribution even when change is
small with $\eta = 1.0$. While the distributions of cmle under known and
estimated cases are also quite identical to each other, there is more
spread in the cmle, with the probability at the true change-point being
substantially smaller than that of the unconditional mle. It is clear
from Figure~\ref{fig1}(b) and (c) that robust to deviations from normality is
quite pronounced even when degrees of freedom under $t_{5}$ and
$\chi_{1}^{2}$ distributions are small. Moving on to $\eta = 2.5$, we
find from Figure~\ref{fig1}(d--f) that, overall, there is greater
robustness and even better agreement between known and estimated
solutions.

Though not presented, the Bias and MSE values show some differences from
known case to the estimated case, mainly when $\eta$ is small ($\eta =
1.0$). The~robustness for large changes ($\eta = 2.5$) is extremely good
throughout the computations, thus depicting good tail behavior for large
changes under both $t$ and $\chi^{2}$ distributions. Also, extreme behavior
is noticed for the estimated case when $\eta = 1.0$ and $n = 100, \tau =
20$. In this case, Cobb's cmle shows somewhat smaller MSE values than
the mle, though only marginally. For all other parameter choices, the
mle performs better in terms of MSE values.

Finally, we noticed that the behavior of MSE values for mle in the known
case are lower than the corresponding theoretical MSE values and that
the MSE values increase with the sample size. This behavior can be
explained by the fact that the theoretical distribution derived for
infinite samples possesses infinite domain, whereas the domain under
finite samples is truncated by the sample size. This truncation effect
for finite samples is found to be most pronounced when $n=40$. The~same
argument also explains why MSE values in both tables increase with
increasing sample sizes.

\section{Multivariate change-point analysis of water discharges at
Nacetinsky creek} \label{sec5}

The~Nacetinsky is a small creek in the German part of the Ergebirge
Mountains. Gombay and Horv\'{a}th (\citeyear{GoHo1997}) analyzed the monthly averages
of water discharges for the Nacetinsky creek during the years 1951--1990
and found that the lognormal distribution appropriately models the
monthly average discharges in the creek. Consequently, applying the log
transformation, they applied likelihood ratio based change detection
methodology in a univariate framework for detecting changes in mean only
as well as changes in the variance only of the normal distribution for
the transformed data. When changes were detected, they obtained point
estimates of the unknown change-point by the value at which the
likelihood ratio was maximum. In detecting the change points, Gombay and
Horv\'{a}th (\citeyear{GoHo1997}) found that the change-detection methodology under
independence was applicable for the monthly water discharges.

We revisited the monthly data and first analyzed the data in a
univariate setup, mainly for detecting changes in mean only or variance
only of the transformed data. Applying the respective likelihood ratio
change-detection statistics (\ref{eqB2}) and~(\ref{eqB4}) in Appendix \hyperref[appendB]{B}, we found no
evidence of change in either the mean or in the variance for almost all
months. We were then interested to learn whether bivariate or
multivariate analyses might convey a different message than what has
been learned from the univariate analysis. One can expect significant
covariances in the water discharges among various months within a year,
and it is of interest to know whether such covariances contribute
significantly as one pursues change-detection and estimation. To this
extent, we found that a multivariate analysis of the data for the months
of February, July and August yields some interesting results.

\begin{figure}

\includegraphics{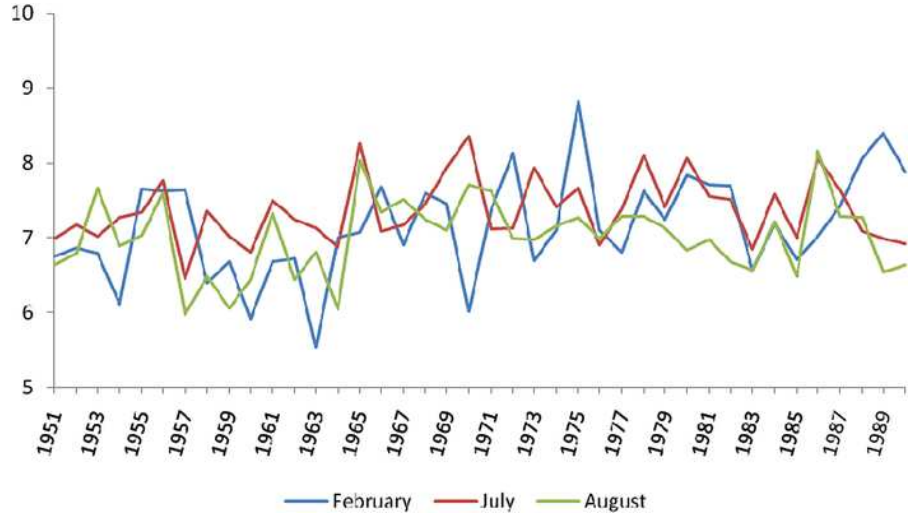}

\caption{Time series plot of log transformed data on mean monthly water
discharges of the Nacetinsky creek for the months of February, July and
August for the years 1951--1990.}\label{fig2}
\end{figure}

Change-point analysis, whether at the univariate level or at the
multivariate level, involves two parts, namely, change-detection and
change-point estimation whenever a change-point is detected. The~focus
of this paper clearly is on estimation, where we derive computable
expressions for the asymptotic distribution of the change-point mle.
Change-detection is not pursued in the theoretical part of this paper.
However, change-detection precedes change-point estimation for the
analysis of data. Keeping this in mind, we first present analysis and
results from change-detection in Appendix \hyperref[appendB]{B}, and only results from
change-point estimation will be emphasized in this section. Once again,
our analysis in both detection and estimation is based on log
transformed water discharges data for the months of February, July and
August as reported in Figure \ref{fig2}.

To proceed with the formulation, let $Y_{i}$ represent the log
transformed monthly water discharges at the Nacetinsky creek for the
months of February, July and August for the for the $i$th year,
$i = 1, \ldots,40$, so that in this case the dimension $d = 3$, and the
sample size $n = 40$. We begin modeling the data by assuming that $Y_{1}, \ldots,Y_{n}$ are independent and that $Y_{i}\sim N( \mu^{( i
)},\Sigma),i = 1, \ldots,n$. Under the change-point setup with
$\tau_{n}$ as the unknown change-point, one lets $\mu^{( i )} = \mu_{1},
i = 1, \ldots,\tau_{n}$ and $\mu^{( i )} = \mu_{2}, i = \tau_{n} + 1, \ldots,n$.

With the above as the basic setup, one can first apply change-detection
methodology, and this has been done comprehensively in Appendix \hyperref[appendB]{B}.
Basically, it has been found that the bivariate tests for Feb--Jul, and
Feb--Aug pairs as well as the multivariate test for all the three months,
were found to be significant even though none of the univariate tests
showed significance. The~bivariate and multivariate analyses resulted in
the change-point mle being $\hat{\tau} _{n} = 14$, so that a change in
water discharges occurred subsequent to the year 1964. The~analysis in
the \hyperref[append]{Appendix} was quite supportive of the assumptions of both Gaussianity
and independence.

We shall now implement the theoretical distribution derived in Section \ref{sec3}
to the data in Figure \ref{fig2} under the bivariate and trivariate cases. Based
on $\hat{\tau} _{n} = 14$, we estimated the values of $\eta$ to
be $\hat{\eta} _{\mathit{FJ}} = 1.47$, $\hat{\eta} _{\mathit{FA}} = 1.52$ and $\hat{\eta}
_{\mathit{FJA}} = 1.60$. Visualizing these as known values, we implemented the
theoretical distribution for each of the three cases. We found the
period 1960--1968 to yield confidence levels of 94.8\%, 95.6\% and
96.5\%, respectively. Simulations suggest that the same period under
both bivariate and trivariate estimated cases with true parameter values
set at $\eta = 1.51$ and $n = 40, \tau = 14$ yields a confidence level
of 90\%. Applying the conditional distribution of Cobb (\citeyear{Co1978}) for the
same data with an error rate of approximately $10^{-5}$, we found that
95\% coverage probability for Feb--Jul is the period 1963--1971, for
Feb--Aug the period is 1963--1969, and for Feb--Jul--Aug the period is
obtained as 1963--1967. Clearly, for this particular data, Cobb's cmle
seems to yield shorter confidence interval than the unconditional mle.
However, under repeated samples for data of the same size with the true
parameters set at $\eta = 1.51$ and $n = 40, \tau = 14$, we found that
the period 1960--1968 under Cobb's cmle yields a coverage probability
of 88\% under both bivariate and trivariate cases, thus showing a
similar performance as the mle on average.

\section{Discussion} \label{sec6}

Asymptotic distribution of the change-point mle is quite complicated and
an exact computable expression for the distribution of the mle has not
been derived in the literature to date, even though Hinkley (\citeyear{Hi1970}, \citeyear{Hi1971},
\citeyear{Hi1972}) published his seminal work more than three decades back. Assuming
the parameters before and after the unknown change-point to be known,
this investigation establishes an exact and yet computationally
attractive form for the asymptotic distribution of the change-point mle,
thus far not available in the literature.

To have a better understanding of its performance, we carried out an
empirical study to compare the distribution under known parameters with
the case where the nuisance parameters remain unknown. We also compare
the derived distribution with the conditional distribution of Cobb
(\citeyear{Co1978}) as well as assessing the robustness of the derived distribution
for departures from normality. Simulations have shown good agreement
between known and estimated cases except for the case where parameters
are estimated and amount of change is relatively small. Also, both mle
and cmle are quite robust to deviations from normality, for the most
part.

We have applied the derived change-point estimation methodology to
compute the asymptotic distribution under both mle and cmle methods for
the log transformed data on annual mean discharges for the months of
February, July and August for the Nacetinsky creek for the years
1951--1990. At first it may appear that sample size of $n=40$ may be
somewhat small for asymptotics to apply. However, simulations under the
estimated case for samples of this size show excellent accuracy in the
univariate case (Table \ref{tab1}, $\eta = 1.5$) and good accuracy in the
bivariate case (Table \ref{tab2}, $\eta = 1.5$). Detection methodology for this
data set under univariate setup yields no significance for the presence
of a change-point for any of the three months. However, change-detection
under the multivariate setup shows significance for Feb--Jul and Feb--Aug
in the bivariate case and also for the trivariate case of Feb--Jul--Aug.

In summary, the methodology proposed in this article appears quite
useful for practitioners in all areas, mainly because it is readily
computable, and it is quite robust to deviations from the assumption
Gaussianity. Also, sample size does not seem to be a serious concern
while implementing the asymptotic result. In terms of future directions,
it would be of interest to derive such computationally feasible
distributions for other distributions such as exponential and Weibull in
the continuous case and binomial and Poisson in the discrete case.

\begin{appendix}\label{append}

\section*{Appendix A}\label{appendA}

\subsection*{Proof of Theorem \protect\ref{teo3.1}}

The~proof of the theorem essentially follows upon applying the following
three lemmas into Theorem \ref{teo2.1}.

The~following lemma is well known [see, e.g., Shiryaev et al. (\citeyear{ShKaKrMe1993})],
and will be given without proof. It should be noted that even though the
original result was given for the continuous Brownian motion, the same
can be applied for a random walk with negative drift. This lemma
addresses the fundamental issue of establishing the distributions of $M$
(and $M^{ *}$) in a simple exponential form, thereby making the
integrals in Theorem \ref{teo2.1} analytically tractable.

\begin{lemma}\label{lem1}
 Let the random walk $\{ S_{n}, n \ge 0 \}$ be as
specified in \textup{(\ref{eq2.3})}. Then, for $x \ge 0$,
\begin{eqnarray*}
P\Bigl( \max_{m \le n}S_{m} \le x \Bigr) &=& \Phi\biggl( \frac{x + n\eta ^{2} / 2}{\sigma
\sqrt{n}} \biggr) - e^{ - x}\Phi\biggl( \frac{ - x + n\eta ^{2} / 2}{\sigma
\sqrt{n}} \biggr) \to 1 - e^{ - x}\\
& =& P( M \le x )\qquad \mbox{as }n \to\infty.
\end{eqnarray*}
\end{lemma}

The~following remark, which provides the complementary probability for
$M$ for strictly positive values ($x > 0$), plays an important role in
the proof of the theorem.

\begin{remark*}
 Note that $P( M \ge x ) = P( M \ge x| M > 0 )P( M > 0 ) = \|
G_{ +} \| e^{ - x},\break x > 0$.
\end{remark*}

The~next lemma provides an analytical and convenient expression for $P(
T_{1}^{ -} > n \cap S_{ - n} \in dx )$. As can be seen from the
proof of Lemma \ref{lem3}, this lemma is critical for carrying out the integrals
in Theorem \ref{teo2.1} in a fully analytical manner.

\begin{lemma}\label{lem2}
 Let the random walk $\{ S_{n}, n \ge 0 \}$ be as
specified in \textup{(\ref{eq2.3})}. Then, for $x \ge 0$,
\begin{eqnarray}
&&P( T_{1}^{ -} > n \cap S_{n} \in dx )= \eta^{ - 1}E\biggl\{ ( T_{1}^{
-} > n - 1 ) \cap\varphi\biggl( \frac{x - S_{n - 1} + \eta ^{2} / 2}{\eta} \biggr)
\biggr\},\nonumber\\
 \eqntext{n \ge 1.}
\end{eqnarray}
\end{lemma}

\begin{pf}
 In light of (\ref{eq3.1}), we have that, for $x > 0$,
\begin{eqnarray}\label{eqA1}
&&P\{ T_{1}^{ -} > n \cap S_{n} \in( 0,x ] \}\hspace*{-8pt}\nonumber\\
&&\qquad = P\Biggl\{ \bigcap_{j = 0}^{n - 1}(
S_{j} > 0 ) \cap S_{n} \in( 0,x ] \Biggr\}\nonumber\\
&&\qquad = E\Biggl[ I\Biggl\{ \bigcap_{j = 0}^{n - 1}( S_{j} > 0 ) \Biggr\}
P\bigl( X_{n} \in( - S_{n - 1},x - S_{n - 1} ] |
\mathsf{F}_{n - 1} \bigr) \Biggr]\nonumber\\
&&\qquad = E\Biggl[ I\Biggl\{ \bigcap_{j = 0}^{n - 1}( S_{j} > 0 )
\Biggr\}\\
&&\quad \qquad {}\times
P\biggl( Z_{n} \in\biggl( \frac{ - S_{n - 1} + \eta ^{2} / 2}{\eta}
,\frac{x - S_{n - 1} + \eta ^{2} / 2}{\eta} \biggr] \Big| \mathsf{F}_{n - 1}
 \biggr) \Biggr]\nonumber\\
&&\qquad = E\biggl[ I( T_{1}^{ -} > n - 1 ) \cap\biggl\{ \Phi\biggl( \frac{x - S_{n - 1} + \eta
^{2} / 2}{\eta} \biggr)\nonumber\\
&&\hspace*{129pt}{} - \Phi\biggl( \frac{ - S_{n - 1} + \eta ^{2} / 2}{\eta} \biggr) \biggr\}
\biggr],\qquad n \ge 1. \nonumber
\end{eqnarray}
Thus, differentiating (\ref{eqA1}) with respect to $x$, the proof of Lemma \ref{lem2} is
now in order.
\end{pf}

The~next lemma provides a manageable expression for the second term in
Theorem \ref{teo2.1}.

\begin{lemma}\label{lem3}
The~following holds:
\[
\int_{ 0 +} ^{ \infty} P( M^{ *} \ge x )P( T_{1}^{ -} > n \cap S_{n}
\in dx ) = \| G_{ +} ^{ *} \|E\{ e^{ - S_{n}}I( T_{1}^{ -} > n )
\},\qquad n \ge 1.
\]
\end{lemma}

\begin{pf}
 Using Lemma \ref{lem2}, and the remark following Lemma \ref{lem1}, we note that
\begin{eqnarray*}
&&\int_{ 0 +} ^{ \infty} P( M^{ *} \ge x )P( T_{1}^{ -} > n \cap S_{n}
\in dx )\\
&&\qquad  = \eta^{ - 1}\| G_{ +} ^{ *} \|E\biggl\{ I( T_{1}^{ -} > n -
1 )\int_{ 0 +} ^{ \infty} e^{ - x}\varphi\biggl( \frac{x - S_{n - 1} + \eta
^{2} / 2}{\eta} \biggr)\,dx \biggr\}\\
&&\qquad  = \| G_{ +} ^{ *} \|E\{ I( T_{1}^{ -} >
n - 1 ) e^{ - S_{n}}I( \eta Z_{n} > - S_{n - 1} + \eta^{2} / 2 ) \}\\
&&\qquad = \| G_{ +} ^{ *} \|E\{ e^{ - S_{n}}I( T_{1}^{ -} > n ) \},\qquad n \ge 1.
\end{eqnarray*}
\upqed\end{pf}

\subsection*{Remarks regarding computational aspects of expressions in
Theorem 3.1}

Here, we first address computational issues of the two
sequences $\{ q_{n}\dvtx n \ge 1 \}$ and $\{ \tilde{q}_{n}\dvtx n \ge 1 \}$ that
appear in Theorem \ref{teo3.1}. Set $b_{n} = P( S_{n} > 0 )$ and $\tilde{b}_{n} =
E\{ e^{ - S_{n}}I( S_{n} > 0 ) \}$, for $n \ge 1$. From Feller (\citeyear{Fe1971}), Volume
II, page 416, and Chover, Ney and Wainger (\citeyear{ChNeWa1973}), it is well known that the
generating function of the sequences $\{ q_{n}\dvtx n \ge 1 \}$ and $\{
\tilde{q}_{n}\dvtx n \ge 1 \}$, respectively, satisfy the following
relationships:
\begin{equation}\label{eqA2}
\sum_{n = 1}^{\infty} s^{n}q_{n} = \exp\Biggl\{ \sum_{n = 1}^{\infty}
\frac{s^{n}b_{n}}{n} \Biggr\}\quad  \mbox{and} \quad \sum_{n = 1}^{\infty} s^{n}\tilde{q}_{n} =
\exp\Biggl\{ \sum_{n = 1}^{\infty} \frac{s^{n}\tilde{b}_{n}}{n} \Biggr\}.
\end{equation}

Note that the second equation in (\ref{eqA2}) appears in Chover, Ney and Wainger (\citeyear{ChNeWa1973})
as a type of a Laplace transform. In addition, both the equations in
(\ref{eqA2}) may be obtained iteratively as simple consequences of the
Weiner--Hopf factorization. In particular, the Leibnitz rule yields the
following iterative relations, and thus enables one to compute $\{
q_{n}\dvtx n \ge 1 \}$ and $\{ \tilde{q}_{n}\dvtx n \ge 1 \}$:
\begin{eqnarray}\label{eqA3}
&&nq_{n} = \sum_{j = 0}^{n - 1} b_{n - j}q_{j}\quad  \mbox{and} \quad n\tilde{q}_{n} =
\sum_{j = 0}^{n - 1} \tilde{b}_{n - j}\tilde{q}_{j},\nonumber\\[-8pt]\\[-8pt]
 \eqntext{n = 1, 2, \ldots,
\mbox{ and }\tilde{q}_{0} = q_{0} = 1.}
\end{eqnarray}

Note that, in the Gaussian case, $b_{n} = \bar{\Phi} ( \eta\sqrt{n} / 2 )
_{}$ and $\tilde{b}_{n} = e^{n\eta ^{2}}\bar{\Phi} ( 3\eta\sqrt{n} / 2
)$, \mbox{$n \ge 1$}.

Next, we demonstrate that the probabilities in Theorem \ref{teo3.1} sum to one,
and then provide an expression for the variance of the limiting
distribution.

From Hinkley (\citeyear{Hi1970}), and the remark after Lemma \ref{lem1} above, it follows that
\begin{eqnarray*}
P( \xi_{\infty} > 0 ) &=& P( M^{*} > M, M^{*} > 0 ) =\int_{ 0 +} ^{
\infty} P( M < x ) P( M^{ *} \in dx )\\
&=& \int_{ 0 +} ^{ \infty} ( 1 - \| G_{ +} \| e^{ - x} ) \| G_{ +} \| e^{
- x}\, dx = 1 - ( 1 - \| G_{ +} \| )^{2} / 2.
\end{eqnarray*}
Since $P( \xi_{\infty} = 0 ) = ( 1 - \| G_{ +} \| )^{2}$, and
$\xi_{\infty}$ is symmetric, the claim that the probabilities for
$\xi_{\infty}$ sum to one follows immediately. The~following expression
for the variance may be derived in a somewhat tedious but
straightforward manner:
\begin{eqnarray*}
\operatorname{Var}( \xi_{\infty} ) &=& 2\{ B''( 1 ) + ( B'( 1 ) )^{2} \}\\
&&{} - 2\exp\bigl( - B( 1
) + \tilde{B}( 1 ) \bigr) \bigl( 1 - \exp( - B( 1 ) ) \bigr) \{ \tilde{B''}( 1 ) + (
\tilde{B'}( 1 ) )^{2} \},
\end{eqnarray*}
where $B( 1 ) = \sum_{ n = 1}^{\infty} b_{n} / n$, $B'( 1 ) = \sum_{ n =
1}^{\infty} b_{n}$, $B''( 1 ) = \sum_{ n = 1}^{\infty} nb_{n}$ and
$\tilde{B}( 1 )$, $\tilde{B'}( 1 )$ and $\tilde{B''}( 1 )$ are defined
upon $\tilde{b}_{n}, n \ge 1$, in a similar manner.

\section*{Appendix B}\label{appendB}

\subsection*{Change-point detection for Nacetinsky water discharges}
 We
first formulate the following hypotheses that test for the presence of
an unknown change-point in the mean vector of the data series:
\renewcommand{\theequation}{B.\arabic{equation}}
\setcounter{equation}{0}
\begin{eqnarray}\label{eqB1}
H_{0}\dvtx  \mu^{( 1 )} &=& \cdots = \mu^{( n )} = \mu_{1}\quad  \mbox{vs.} \nonumber\\[-8pt]\\[-8pt]
H_{a}\dvtx  \mu^{(1 )} &=& \cdots = \mu^{( \tau )} = \mu_{1} \ne\mu^{( \tau + 1 )} = \cdots
= \mu^{( n )} = \mu_{2},\nonumber
\end{eqnarray}
where $\tau\in\{ 1, \ldots,n - 1 \}$ is the unknown
change-point. Asymptotic theory of the generalized likelihood ratio
statistic for testing the above hypothesis has been well addressed in
the literature and the limiting result may be found in Cs\"{o}rg\H{o}
and Horv\'{a}th (\citeyear{CsHo1997}). It may be shown that the twice log-likelihood
ratio statistic for testing the above hypothesis is
\begin{equation}
\label{eqB2}
U_{n} = \max_{1 \le t \le n - 1}n\log( | \hat{\bolds{\Sigma}}
_{n} | / | \hat{\bolds{\Sigma}} _{t} | ),
\end{equation}
where $\hat{\bolds{\Sigma}} _{t} = n^{ - 1}\{ \sum_{i = 1}^{t} (
\mathbf{Y}_{i} - \hat{\bolds{\mu}} _{1,t} ) ( \mathbf{Y}_{i} -
\hat{\bolds{\mu}} _{1,t} )^{T} + \sum_{i = t + 1}^{n} ( \mathbf{Y}_{i}
- \hat{\bolds{\mu}} _{2,t} ) ( \mathbf{Y}_{i} - \hat{\bolds{\mu}}
_{2,t} )^{T} \}$, $\hat{\bolds{\mu}} _{1,t} = t^{ - 1}\sum_{i = 1}^{t}
\mathbf{Y}_{i}$ and $\hat{\bolds{\mu}} _{2,t} = ( n - t )^{ - 1}\sum_{i
= t + 1}^{n} \mathbf{Y}_{i}, t = 1, \ldots,n$. The asymptotic
distribution of the above statistic is based upon $W_{n} =\break ( 2\log\log
nU_{n} )^{1 / 2} - ( 2\log\log n + \frac{p}{2}\log\log\log n -
\log\Gamma( p / 2 ) )$, where $p$ denotes the number of parameters that
change under the alternative hypothesis, and in this case we have $p = d
= 3$. The~limiting distribution of $W_{n}$ is given by the following
double exponential form:
\begin{equation}
\label{eqB3}
\lim_{n \to \infty} P[ W_{n} \le t ] = \exp( - 2 e^{ -
t} ).
\end{equation}

The~$p$-value is obtained based on a two-sided critical region of the
above limiting distribution. When a test is significant, the maximum
likelihood estimator of the unknown change-point $\tau$ is obtained as
the argument at which $U_{n}$ attains its maximum. In principle, we may
apply the above procedure for the data of each month individually with $p
= 1$, and also for data on each pair of months with $p = 2$. The~results
of the tests for all cases are presented in Table \ref{tab3}. Clearly, all
univariate tests are not significant. Among the bivariate tests, the
pair July--August is not significant, whereas the other two pairs yield
significance. The~multivariate test for all three months is also
significant. The~significance based upon the biviariate and multivariate
tests takes into account the covariance structure in the data and hence
should be believed more so than the univariate tests where no
significance is found. The~change-point mle is obtained as $\hat{\tau} =
14$.

\makeatletter
\long\def\tabnotes#1{\vskip 3pt
\begin{minipage}
[t]{12cm}#1
\end{minipage}}
\makeatother
\begin{table}[b]
\tablewidth=8cm
\caption{The~statistic $W$ for change in mean for various months and their $p$-values}
\label{tab3}
\begin{tabular*}{8cm}{@{\extracolsep{\fill}}lccc@{}}
\hline
\textbf{Months} & $\bolds{W}$ & $\bolds{p}$\textbf{-value} & $\bolds{\hat{\tau}}$ \\
\hline
Feb & 2.74 & 0.1206 & 15\\
Jul & 1.86 & 0.2674 & 14\\
Aug & 2.29 & 0.1825 & 14\\
Feb--Jul & 3.59 & 0.0539 & 14\\
Feb--Aug & 3.76 & 0.0455 & 14\\
Jul--Aug & 1.90 & 0.2593 & 14\\
Feb--Jul--Aug & 3.78 & 0.0448 & 14\\
\hline
\end{tabular*}
\end{table}

\begin{table}[b]
\tablewidth=8cm
\caption{The~statistic $W$ for change in variance for various months and their $p$-values}
\label{tab4}
\begin{tabular*}{8cm}{@{\extracolsep{\fill}}lccc@{}}
\hline
\textbf{Months} & $\bolds{W}$ & $\bolds{p}$\textbf{-value} & $\bolds{\hat{\tau}}$ \\
\hline
Feb & 3.18 & 0.0796 & 3\\
Jul & 1.91 & 0.2556 & 5\\
Aug & 1.39 & 0.3929 & 2\\
Feb--Jul & 3.02 & 0.0927 & 3\\
Feb--Aug & 2.28 & 0.1842 & 2\\
Jul--Aug & 2.32 & 0.1788 & 2\\
Feb--Jul--Aug & 4.26 & 0.0278 & 3\\
\hline
\end{tabular*}
\end{table}

At this point, we need to investigate the validity of the main
assumptions, namely, constancy of the covariance matrix, Gaussianity and
independence over time. The~investigation regarding the covariance
matrix requires that we compute the deviation vector $D_{i}, i = 1, \ldots,40$, from the estimated mean for each observation, taking into
account the differences in the means before and after the estimated
change-point. It is of interest then to know whether the covariance
structure of the deviations remained constant throughout the sampling
period. The~generalized log-likelihood ratio statistic for the constancy
of the covariance matrix over time against the alternative that the
covariance matrix has changed at an unknown time is given by
\begin{equation}
\label{eqB4}
U_{n}^{ *} = \max_{1 \le t \le n - 1}\log\bigl\{ |
\hat{\bolds{\Sigma}} _{1\dvtx n} |^{n} / \bigl( |
\hat{\bolds{\Sigma}} _{1\dvtx t} |^{t} | \hat{\bolds{\Sigma}}
_{t + 1\dvtx n} |^{( n - t )} \bigr) \bigr\},
\end{equation}
where $| \hat{\bolds{\Sigma}} _{1\dvtx t} |$ and $| \hat{\bolds{\Sigma}}
_{t + 1\dvtx n} |$ are the usual estimators of the covariance matrix based on
the first $t$ and last $n - t$ deviations, respectively. The~limiting
distribution of $U_{n}^{ *}$ is obtained through the distribution of
$W_{n}^{ *}$, where $W_{n}^{ *}$ is defined upon $U_{n}^{ *}$ in an
analogous manner. It follows that $p$, the number of parameters that
change in this case, is given by $p = d( d + 1 ) / 2$. The~$p$-values for
the univariate, bivariate and multivariate tests are reported in Table
\ref{tab4}. Clearly, all tests are insignificant except the multivariate test.
However, the significance is not particularly relevant since the
change-point mle of 3 obtained in this case implies no change in the
covariance structure, for all practical purposes. Thus, there is no
evidence in the data against the assumption of stationarity of the
covariance matrix. Utilizing the estimated change-point ($\hat{\tau} =
14$), estimates for the mean vector before and after the change-point as
well as the pooled estimator of the common covariance matrix are then
obtained as $\hat{\mu} _{1\hat{\tau}} = ( 6.738, 7.137, 6.725 ),
\hat{\mu} _{2\hat{\tau}} = ( 7.383, 7.483, 7.166 )$ and
\[
\hat{\Sigma} _{\hat{\tau}} = \left[ \matrix{
 0.365 & - 0.032 & -0.029 \cr
  - 0.032 & 0.161 & 0.104 \cr
   - 0.029 & 0.104 & 0.211}\right].
   \]
It remains to be seen whether the assumptions of Gaussianity and
independence over time are valid. We can verify this by utilizing the
deviation vectors $D_{i}$, \mbox{$i = 1, \ldots,40$}, and the covariance matrix
$\hat{\Sigma} _{\hat{\tau}}$ found above. Specifically, if $D_{i}$ is
multivariate normal, then it is well known that $d_{i}^{2} = \| D_{i}
\|_{\hat{\Sigma} _{\hat{\tau}} ^{ - 1}}^{2}$ is approximately chi-square
with 3 degrees of freedom $i = 1, \ldots,40$. The~same can be applied for
the bivariate case also with the degrees of freedom being 2 in this
case. Thus, one only needs to verify whether $d_{i}^{2}, i = 1, \ldots,40$ form a sample from the corresponding chi-square distribution.
Upon applying the Anderson--Darling statistic, we found the $p$-value for
the three months case to be 0.185. The~corresponding $p$-values for
Feb--Jul, Feb--Aug and Jul--Aug pairs were 0.244, 0.250 and 0.10,
respectively. In the univariate case, we applied the Anderson--Darling
test for the deviations for each individual month and found the $p$-values
to be 0.927, 0.530 and~0.177, respectively. Thus, the assumption of
Gaussianity seems quite appropriate at each of the univariate, bivariate
and multivariate levels.

As for independence over time, we first tested each of the three
deviation series for significance of both autocorrelations and partial
autocorrelations up to the first twenty lags. The~ACF and PACF plots for
each individual series showed no evidence of significant correlations.
We then computed the cross-correlations for each pair and found that
these were also not significant and, thus, there was no indication that
the assumption of independence over time was in violation. Overall, the
change-point model with estimated parameters may be seen to fit the data
quite well.
\end{appendix}

\section*{Acknowledgments}
 The~authors thank the Editor Michael Stein, the
Associate Editor and two anonymous referees for their in-depth comments
and suggestions that led to a substantial improvement in both content
and presentation of the paper. We are especially thankful to Professor
Daniela Jaru\v{s}kov\'{a} for providing us the data on Nacetinsky creek.

\printaddresses

\end{document}